**Impact of Solar Particle Events on Space Radiation Shielding: OLTARIS Simulation and Quantum Optimization of Material Selection using QAOA and VQE Algorithms**

**Author (s): Dr. Kavita Lalwani[1], Sreedevi V. V.[1], Munish Gakhar[2], Diya Agrawal[3]**

[1] Malaviya National Institute of Technology Jaipur,

[2] IISER Pune,

[3] ARSD, University of Delhi

**ABSTRACT**

Space radiation poses a significant challenge for long-duration human space missions, with sources including Galactic Cosmic Rays (GCRs), Solar Particle Events (SPEs), and trapped particles in regions of Van Allen belts. These high-energy radiations can cause severe biological effects on astronauts and degrade spacecraft systems, making effective shielding a critical requirement. Traditionally, passive shielding materials such as aluminum have been used, but their limitations, especially in generating secondary radiation, necessitate the search for more efficient alternatives. In this study, the performance of various shielding materials such as Lithium hydride (LiH), Polyethylene ($C_2H_4$), Lithium borohydride ($LiBH_4$), Beryllium borohydride ($Be[BH_4]_2$) and Ammonia borane ($NH_3BH_3$) are systematically evaluated in the GCR & SPE environment using OLTARIS, a NASA-developed simulation tool. Further, the effect of solar particle events (October 1989) on the performance of shielding materials are studied and evaluated through particle flux distributions (energy spectrum) and equivalent dose distributions for these materials. The relative shielding effectiveness of these materials in the SPE environment varied from that in the GCR environment suggesting that material performance is dependent on the radiation environment. Beryllium borohydride emerges as the best material in the SPE environment while lithium hydride gives the least dose equivalent in the GCR environment. The particle wise dose equivalent and flux contributions reveal that, in the SPE environment, the material performance is directly linked to the hydrogen content of the material. In addition, the effect of solar modulation on dose equivalent in the GCR environment is studied. As solar modulation increases, the GCR intensity reduces and hence dose equivalent also reduces.

As the space radiation environment consists of complex particle radiation of high energies, in order to validate the large sample of materials for space radiation shielding, the combinatorial complexity of possible configurations presents a major computational challenge. To address this, the material selection problem was formulated as a Quadratic Unconstrained Binary Optimization (QUBO) model and solved using the Variational Quantum Eigensolver (VQE) and Quantum Approximate Optimization Algorithm (QAOA). By mapping OLTARIS dose-equivalent data to the Ising model and utilizing the quantum-classical hybrid approaches of VQE and QAOA, optimal shielding configurations that minimize radiation dose were identified, demonstrating the potential of quantum algorithms in advancing space radiation protection strategies. Our results are in agreement between OLTARIS and Quantum Optimization in both the environments (GCR, SPE).

**KEY WORDS:** GCR, SPE, OLTARIS, Flux, Dose equivalent, VQE, QAOA, QUBO, IBM, Ising Model, Hamiltonian, Ansatz, Quantum Circuit.

## 1. INTRODUCTION

As technology advances, the pursuit of human civilization beyond Earth and the increase in manned space missions are becoming more prominent. However, space is far from tranquil. One of the major challenges for humans in space is exposure to radiation from the Sun and other stars, both within and beyond our galaxy [1]. Intense fluxes of high-energy radiation (ranging from a few MeVs to ~1 TeV) can cause temporary cellular damage, blood abnormalities, nausea, fatigue, reduced white blood cell counts, radiation sickness, vomiting, and, in extreme cases, death. On Earth, we are shielded from these effects by its own magnetic field, which acts as a natural barrier against harmful radiation. In space, however, this protective shield is absent. As a result, careful consideration of radiation exposure is essential when planning for deep space missions. These radiation sources not only pose serious health risks to astronauts but also threaten the performance and durability of spacecraft electronics and materials. Consequently, understanding and mitigating space radiation is critical to ensuring the safety and success of deep space exploration.

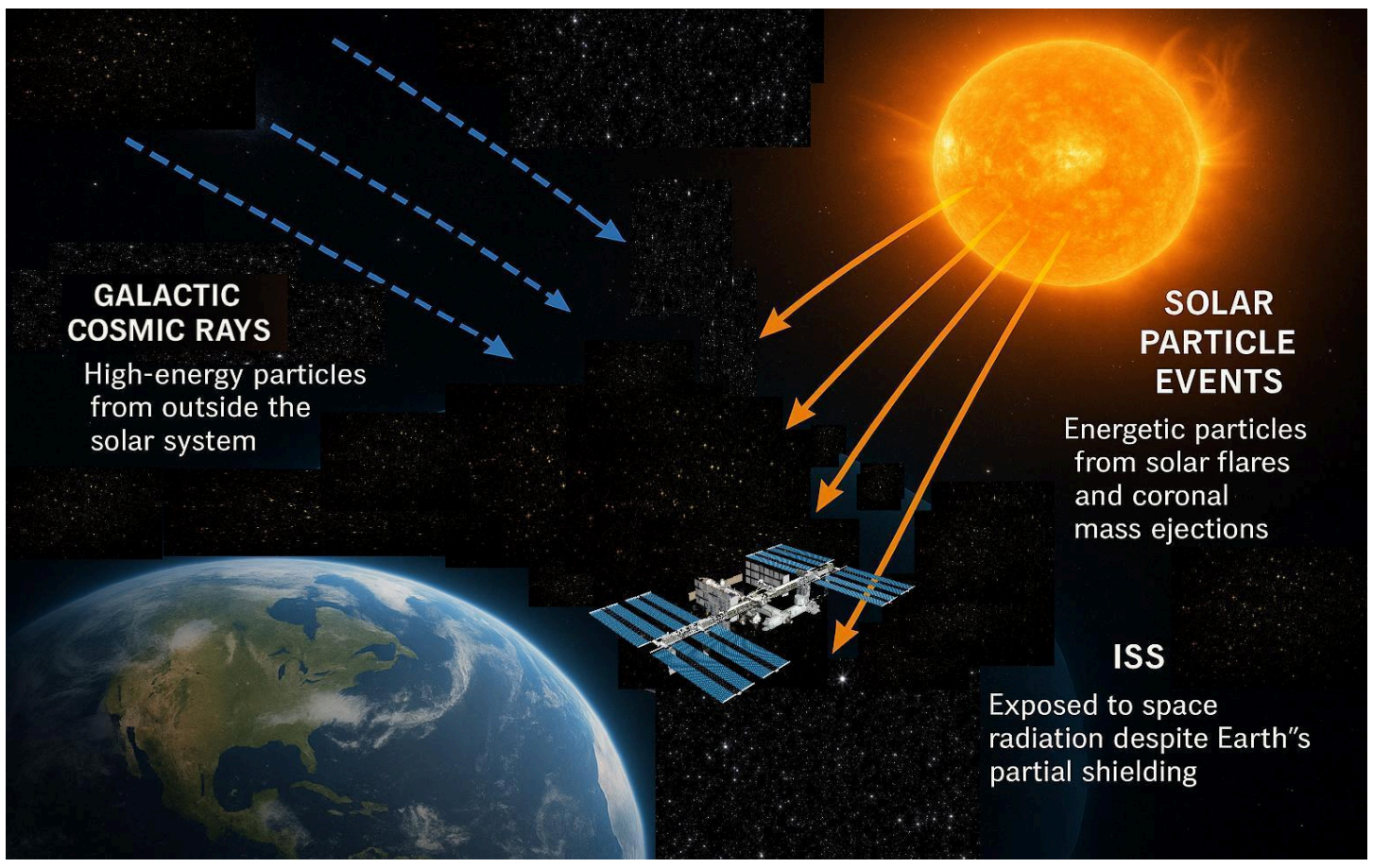

***Fig. 1**: Space Radiation Environment.*

There are three main sources of space radiation: (i) Galactic Cosmic Rays (GCRs) [2], (ii) Solar Particle Events (SPEs) [3], and (iii) Ionized particles confined within regions such as the Van Allen radiation belts. For missions far beyond Earth's orbit, GCRs pose the greatest risk. These radiations are made up of about 85% protons, 13% alpha particles, and a small percentage of heavier nuclei from elements like carbon and iron. Both GCRs and SPEs are harmful, but GCRs are incredibly energetic and can easily pass through spacecraft walls, sometimes creating even more radiation in the form of secondary particles when they interact with shielding materials. On the other hand, SPEs are more unpredictable and short-lived, while trapped radiation is mostly a concern for satellites or astronauts in Low Earth Orbit (LEO). Therefore, adequate radiation shielding should be an essential component while long-duration, manned space missions are planned. To implement effective radiation shielding strategies, it is essential to first understand the fundamental principles behind radiation protection. There are two ways in which radiation shielding can be implemented:

active shielding and passive shielding [4]. Active shielding uses magnetic and electric fields, but this method has its own advantages and disadvantages. Advantages are that the setup would be mass efficient, better against SPEs, reusable, and more controllable. Some of the disadvantages are that this would require a high power supply and complex structures. Passive shields are material layer shielding, a material that is placed between a source of radiation and tissue to absorb radiation falling on it. Traditionally, Aluminum-based materials are used in space exploration missions as a shielding layer, but they are not ideal, as Aluminum produces high amounts of secondary radiations, and hence, dose and flux will be significantly higher than the acceptable limits. Researchers all over the world are trying to figure out a much more efficient shielding material that not only reduces dose and flux but is also feasible in space missions for different applications. In this study, the performance of various types of shielding materials in the GCR environment is evaluated using OLTARIS [5]. Our primary focus is to investigate the effect of Galactic Cosmic Rays and Solar Particle Events on the performance of shielding materials and to identify the most effective shielding material by optimizing their response within this space radiation environment (GCR & SPE).

Due to the intricate nature of high-energy particle radiation in space, assessing a broad range of materials for radiation shielding poses a major computational challenge stemming from the combinatorial complexity of configuration possibilities. Quantum Optimization offers a promising framework to address such complex material selection tasks. As the problem size increases especially when using OLTARIS to evaluate a larger number of shielding materials and their possible configurations, the combinatorial space grows exponentially. For example, simulating different materials with multiple thicknesses or layering options for each leads to a vast number of possible shielding configurations. This rapid expansion in complexity makes it increasingly difficult for classical algorithms to determine exact solutions within practical timeframes, as they must exhaustively evaluate a massive number of combinations. To address this challenge, the material selection task is framed as a Quadratic Unconstrained Binary Optimization (QUBO) problem and solved using hybrid quantum-classical methods, specifically the Variational Quantum Eigensolver (VQE) and the Quantum Approximate Optimization Algorithm (QAOA) implemented on a Quantum Simulator (IBM Aer Simulator). In this work, we investigate the application of the Variational Quantum Eigensolver (VQE) [6] and the Quantum Approximate Optimization Algorithm (QAOA) [7] to Quadratic Unconstrained Binary Optimization (QUBO) [8] problems mapped onto the Ising model, utilizing the IBM AerSimulator [9]. Our study focuses on a material selection problem for space radiation shielding, employing dose-equivalent data sourced from OLTARIS. The primary objective is to minimize the equivalent radiation dose across varying shielding depths. Although VQE was initially developed for quantum chemistry, both VQE and QAOA have demonstrated significant adaptability in solving optimization problems formulated within the QUBO framework [10]. In our methodology, the shielding material selection is formulated as a QUBO instance corresponding to each shielding depth. These instances are then mapped onto the Ising Hamiltonian and solved using VQE and QAOA to approximate their ground states. The resulting configurations represent the optimal material choices that minimize radiation dose, thereby enhancing the effectiveness of space radiation shielding.

## 2. MATERIALS AND SIMULATION SETUP

This section presents a detailed discussion of the radiation shielding materials selected for the study, followed by an introduction to the simulation methodology employed using OLTARIS. In addition, a quantum optimization setup using IBM's Aer Simulator is introduced, where the material selection problem for optimizing the space radiation shielding is formulated as a QUBO and solved using VQE and QAOA algorithms.

### 2.1 Radiation Shielding Materials

In the context of space radiation, the selection of shielding materials is highly constrained. An ideal shielding material must not only reduce the biological dose effectively but also exhibit chemical and mechanical stability under extreme environmental conditions. Key desirable properties include high hydrogen content, low secondary particle production, moderate atomic-to-mass ratio (A/Z), chemical inertness, and effective neutron absorption. High-density materials provide robust structural support but often lead to increased secondary particle production, thereby increasing the radiation dose. Conversely, low-density, hydrogen-rich materials generally result in reduced secondary radiation, but may lack structural strength and exhibit higher chemical reactivity. Conventionally, Aluminum is widely utilized in spacecraft shielding due to its favorable structural properties. However, investigations reveal that Aluminum generates significant secondary particle radiations, prompting the search for alternative materials with superior shielding performance. Among the most effective shielding substances identified is hydrogen, which provides minimal radiation dose. Nevertheless, practical implementation of hydrogen in space applications remains a challenge due to its physical state and containment issues. Hydrocarbons like Polyethylene are of great interest in this regard as they have high hydrogen content as well as required structural integrity. Metal hydrides [11-13] such as Lithium hydride also demonstrate promise owing to their low atomic number, high hydrogen content, and moderate A/Z ratio.

In this study, the materials selected for evaluation include Lithium hydride (a representative metal hydride), Polyethylene (a hydrogen-rich polymer), and Aluminum (as a conventional reference material). Table 1 lists the materials considered, along with their chemical formulas and respective densities. Materials in table 1 are arranged in the increasing order of their hydrogen content.

| **S.N.** | **Material name** | **Chemical formula** | **Density (g/cm²)** |
|---|---|---|---|
| 1 | Aluminum | Al | 2.7 |
| 2 | Lithium hydride | LiH | 0.82 |
| 3 | Polyethylene | $C_2H_4$ | 0.94 |
| 4 | Lithium borohydride | $LiBH_4$ | 0.66 |
| 5 | Ammonia borane | $NH_3BH_3$ | 0.78 |
| 6 | Beryllium borohydride | $Be[BH_4]_2$ | 0.604 |

*Table 1: Materials used for space radiation shielding in this study, their chemical formula, and their corresponding density.*

To simulate the materials as listed in table 1 for space radiation shielding, OLTARIS simulations are employed and are discussed in section 2.2. To further optimize the various shielding materials in the complex space radiation environments (GCR, SPE), Quantum Optimization Framework is used and compared with the Classical optimization method and is discussed in section 2.3.

### 2.2 OLTARIS Simulation Framework

OLTARIS (On-Line Tool for the Assessment of Radiation in Space) is a web-based platform developed by NASA for assessing radiation transport and shielding effectiveness in space environments. OLTARIS is an advanced iteration of NASA's earlier SIREST (Space Ionizing Radiation Effects and Shielding Tools) system and is developed through user-driven enhancements and modern computational design practices.

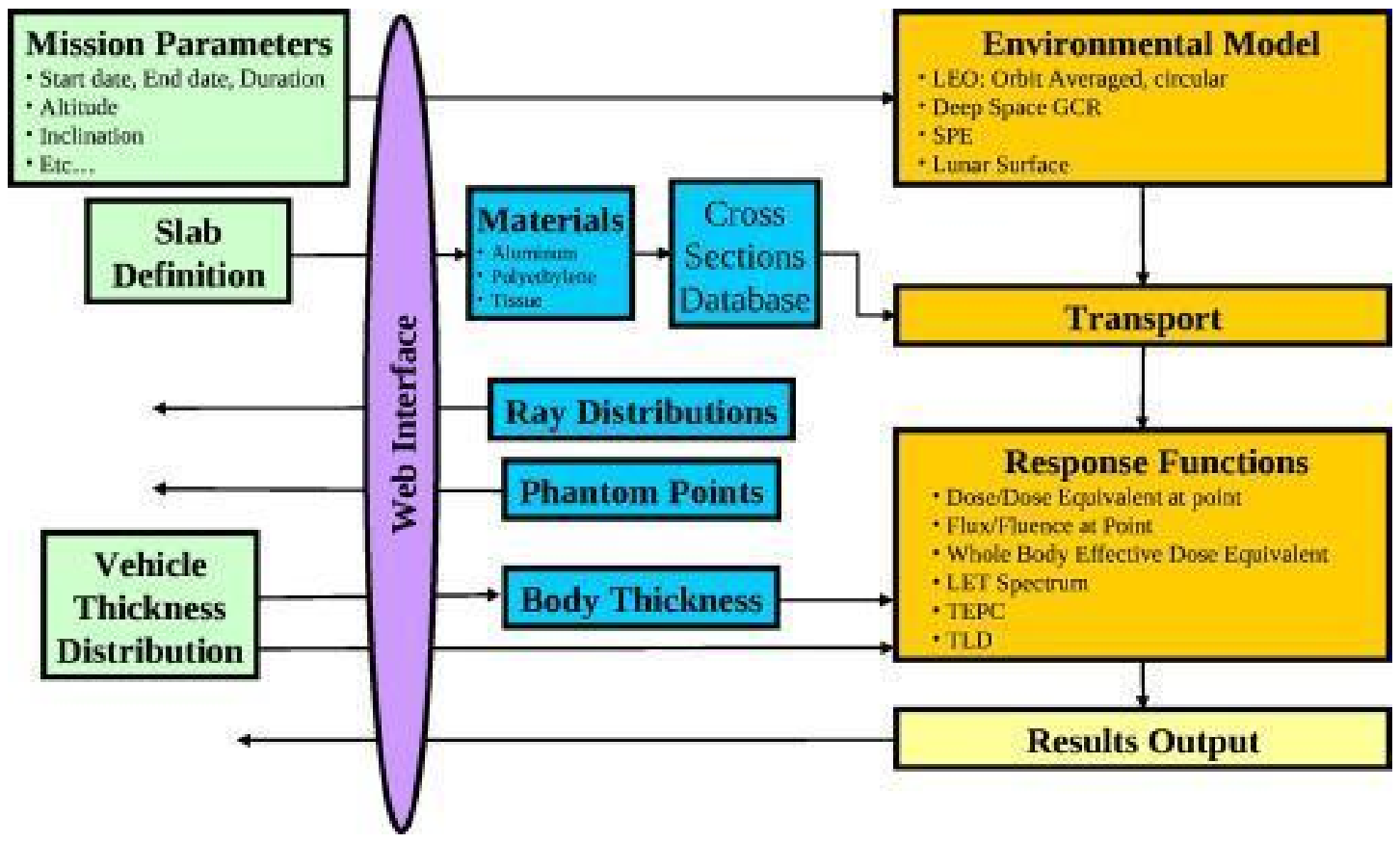

***Fig. 2****. Flowchart of simulation of shielding materials in OLTARIS.*

Fig. 2 presents a schematic mind map of the OLTARIS framework. Input parameters (shown in green) include mission details such as start and end dates, mission duration, and shielding geometry (e.g., slab and spherical configurations). The user interface (purple) facilitates interaction with the platform. Material parameters such as composition, density, and thickness (blue) are specified along with the desired radiation environment and expected outputs (yellow), such as Galactic Cosmic Rays (GCR) in free space, Solar Particle Events (SPE) on lunar surfaces, and the desired response quantities like flux or dose. OLTARIS utilizes the Boltzmann Transport Equation (BTE) to model particle transport and interactions in the space radiation environment and is expressed by equation (1).

$$\frac{\partial \psi}{\partial t} + \mathbf{v} \cdot \nabla \psi + \Sigma_t \psi = \int \Sigma_s(\mathbf{r}, E' \to E)\, \psi(E')\, dE' \ldots (1)$$

$\frac{\partial \psi}{\partial t}$ : Time change of particle flux, $v \cdot \nabla\psi$: Particles streaming through space, $\sum_t \psi$ : Loss due to interactions of particles with materials (attenuation), $\int \sum_s (r,\ E' \to E)\psi(E')dE'$: Gain from scattering into energy E.

The equation (1) accounts for temporal changes in particle flux, streaming effects, attenuation due to interactions, and contributions from scattering. It forms the mathematical backbone of OLTARIS and enables accurate simulation of particle behavior in shielding materials. It incorporates various particle interactions and are listed in section 2.2

**2.2.1 Interactions of Space Radiation with Materials:**
**Ionization and Excitation (via the Bethe-Bloch Equation)**
Charged particles (protons & alpha particles) in the space radiation environment, lose their energy as they pass through material, primarily by ionizing atoms or exciting electrons. This energy loss is described by the Bethe-Bloch equation, which estimates the stopping power based on properties of the particle (charge, velocity) and the material (atomic number, mean excitation potential, etc.).

$$S = -\frac{dE}{dx} = Kz^2 \cdot \frac{Z}{A} \cdot \frac{1}{\beta^2} \left[ \ln\left( \frac{2m_e c^2 \beta^2 \gamma^2 T_{\max}}{I^2} \right) - \beta^2 - \frac{\delta(\beta\gamma)}{2} \right] \ldots (2)$$

$S = -\frac{dE}{dx}$: Stopping power, $e$: Elementary charge, $m_e$: Electron mass, $c$: Speed of light, $Z$: Atomic number of the absorber, $A$: Mass number of the absorber, $T_{max}$: Maximum energy transferable to an electron in a single collision, $\delta(\beta\gamma)$: Density effect correction to ionisation loss, $I$: Mean excitation potential of the absorbing medium.

Using this equation (2), we can estimate energy loss and radiation dose by particle radiations which are interacting with materials in the space radiation environment.

**Nuclear Fragmentation:**Heavy ions such as iron (Fe-56) present in GCR can undergo nuclear interactions with shielding nuclei, resulting in fragmentation and the generation of secondary particles. These secondary particles contribute to the overall radiation dose. In OLTARIS, this process is modeled using the HZETRN (High-Z and Energy Transport) code [14].

**Elastic and Inelastic Scattering:** These processes involve the deflection of incoming particles by nuclei. In elastic scattering, the incoming particle collides with a nucleus and bounces off without changing the internal state of the nucleus, conserving kinetic energy. In inelastic scattering, part of the kinetic energy is absorbed by the nucleus, exciting it and potentially leading to emission of secondary particles like gamma rays or neutrons.

In this study, the shielding effectiveness of selected materials is evaluated using slab geometry with an areal density of 15 g/cm². This implies that the product of the physical thickness and density of the material equals 15 g/cm². Since OLTARIS calculates the response functions after each layer, each shielding configuration consists of 15 layers. This is helpful in evaluating dose equivalent and radiation flux within the depth of the shield. Simulations were conducted for a 1-day mission duration in free space at 1 AU from the Sun. The radiation response is evaluated in tissue using quality factors defined in ICRP 60 [15].

The schematic representation of the simulation set up is shown in figure 3. The figure shows how a hydrogen rich material effectively attenuates the radiation from the sun and allows a minimum amount to pass through. The dose equivalent is calculated from the energy deposited in the human tissue by the radiation after passing through the shield.

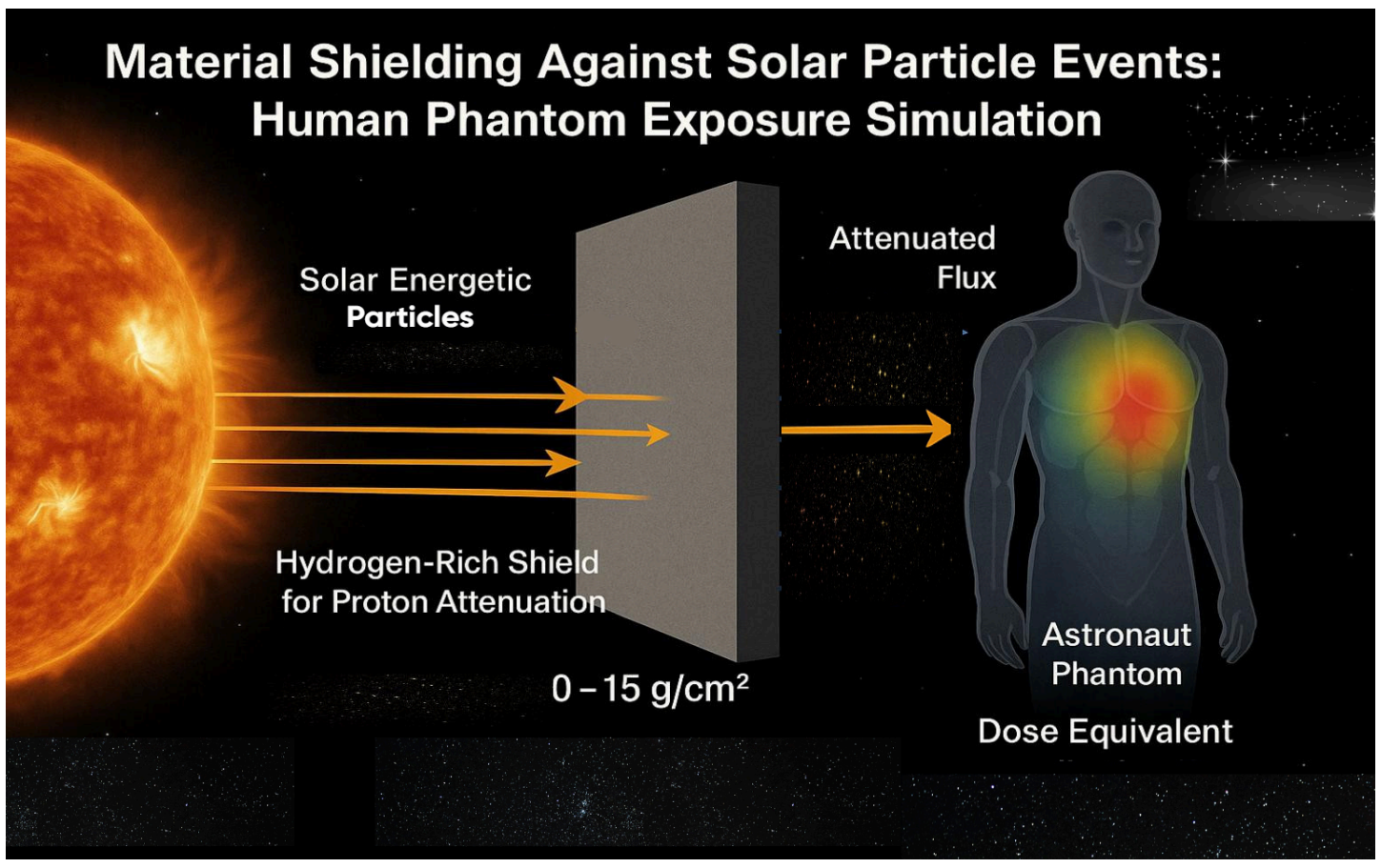

***Fig.3****: Material shielding against SPE.*

### 2.3 Material optimization using Quantum Algorithms on Quantum Simulator

Given the complex nature of high-energy particle radiations in the space environment, validating a large set of materials for space radiation shielding involves a significant computational challenge due to the combinatorial complexity of possible configurations. To tackle this issue, we represent the material selection problem as a Quadratic Unconstrained Binary Optimization (QUBO) instance and solve it using Quantum Optimization techniques, specifically the Variational Quantum Eigensolver (VQE) and the Quantum Approximate Optimization Algorithm (QAOA) on a Quantum Simulator (IBM Aer Simulator) [9].
Figure 4 shows the flowchart, which illustrates the VQE & QAOA-based QUBO optimization workflow using Aer Simulator. In the following sub-sections, details of each step given in the flowchart are explained.

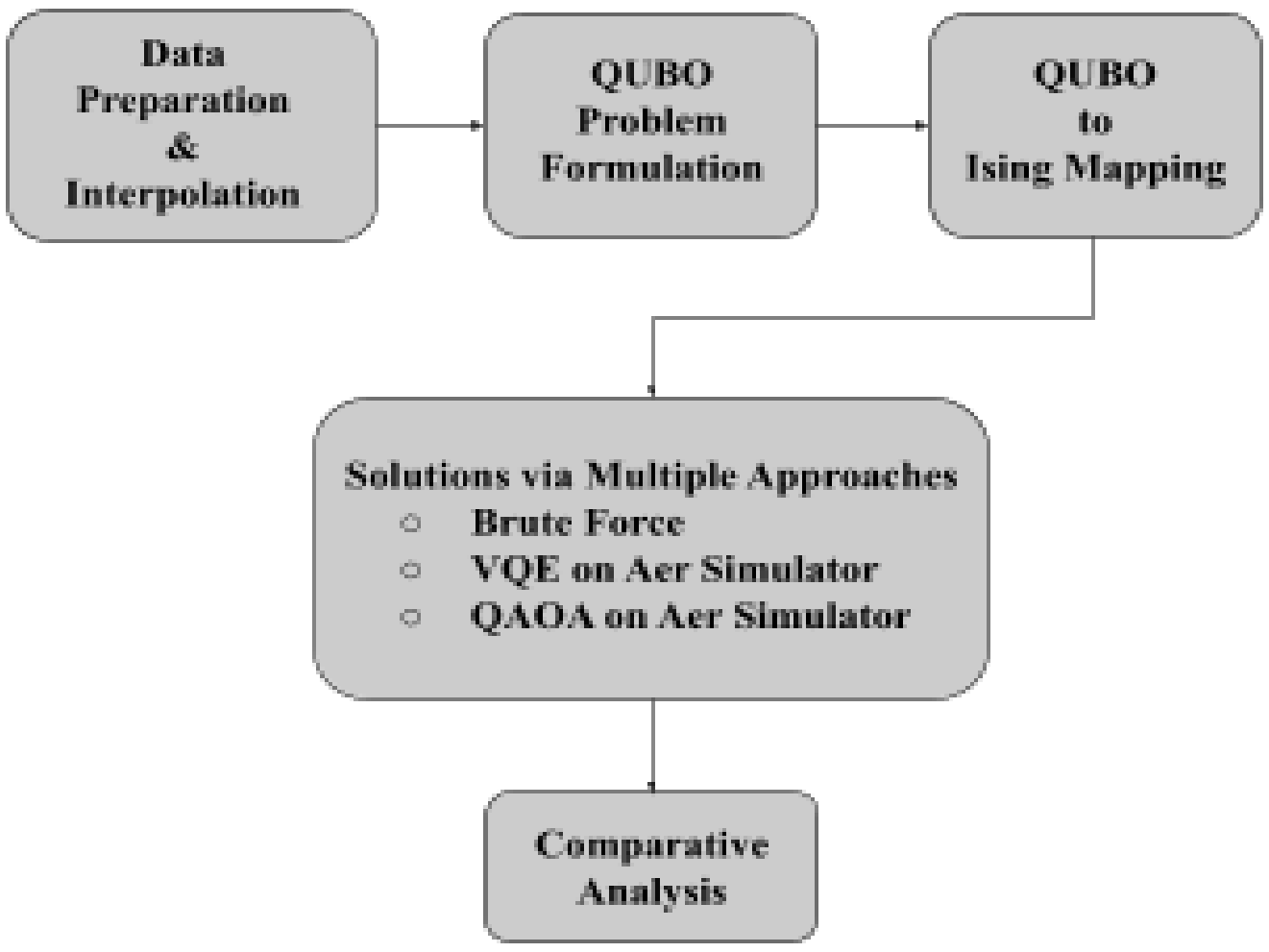

***Fig. 4*** *Flowchart illustrating the VQE & QAOA-based QUBO optimization workflow*

### 2.3.1 Data Preparation & Interpolation

To select the best material with minimum dose using quantum simulation, the dose equivalent data is generated from OLTARIS simulations and fed into the optimization model. Due to non-uniform spatial sampling of depth profiles across materials (Aluminum, Polyethylene, Lithium Hydride, Lithium borohydride, Ammonia borane, and Beryllium borohydride), the data is carefully interpolated onto a common set of depth points, ensuring the original values remain unaltered. As we are employing variational algorithms to select the material with minimum dose, the dose values are globally normalized within a 0–100 range, which will help us to guide the method toward the correct minimized solution.

### 2.3.2 Quadratic Unconstrained Binary Optimization (QUBO Model)

The normalised output obtained from section 2.3.1 is encoded into a QUBO model [8] with the objective of minimizing the quadratic function, which is expressed by equation (3). Here, the QUBO model is a standard mathematical representation for combinatorial optimization problems involving binary variables.

$$f(x) = x^T Q x + c^T x + d \ \ldots (3)$$

where $x \in \{0, 1\}^N$, $Q \in \mathbb{R}^{N\times N}$ is a symmetric matrix of quadratic coefficients, $c$ is a vector of linear coefficients, and $d$ is a constant [16].

In this study, each variable $x_{i,j}$ represents the selection of material $j$ at depth $i$, with $x_{i,j} = 1$ if the material is selected, and 0 otherwise. The goal is to select one material per depth such that the total absorbed radiation dose is minimized in both GCR and SPE.
To ensure exactly one material is selected per depth, a quadratic penalty term is added to the objective, and is given by equation (4):

$$\lambda \sum_{j=1}^{M} (x_{i,j} - 1)^2 \ldots (4)$$

where $\lambda$ is a tunable penalty parameter and $M$ is the number of materials (three materials for GCR and six materials for SPE). A dynamic penalty approximately 10 times the maximum dose per depth is set, which enforces valid configurations within the QUBO formulation by heavily penalizing multiple or no selections.

### 2.3.3 Mapping QUBO to Ising Hamiltonian

In order to get the bitstring (0 when material is not selected or 1 when material is selected) from equation (4), we encoded the QUBO model into an Ising Hamiltonian. The Ising model is widely used in optimization by mapping problem variables to spin states [17], enabling solution search through energy minimization, and it is expressed by equation (5).

$$H(s) = \sum_i h_i s_i + \sum_{i<j} J_{ij} s_i s_j \ \ldots (5)$$

Quantum optimization algorithms typically operate on spin variables $s_i \in \{-1, +1\}$. To map the QUBO to an Ising form, we use the transformation that is defined by equation (6).

$$x_i = \frac{1 + s_i}{2} \Rightarrow s_i = 2x_i - 1 \;\ldots (6)$$

where $h_i$ are local field coefficients and $J_{i,j}$ are interaction strengths between spins [18]. Each Hamiltonian acts on three & six qubits, directly corresponding to the three & six candidate materials available at that depth for GCR and SPE, respectively. This direct mapping obviates the need for qubit reuse, encoding, or ancillary variables, simplifying the quantum circuit design and ensuring one-to-one correspondence between physical qubits and decision variables.

### 2.3.4 Variational Quantum Eigensolver (VQE)

After mapping the material selection problem into an Ising Hamiltonian (Section 2.3.3), we employ the Variational Quantum Eigensolver (VQE) [6] to approximate its ground state. Since the ground state of the Ising Hamiltonian corresponds to the optimal configuration that minimizes the total dose, VQE serves as a suitable approach for solving the encoded optimization problem of space radiation shielding. VQE leverages the variational principle of quantum mechanics, which asserts that for any parameterized quantum state $|\psi(\vec{\theta})\rangle$, the expectation value of a Hamiltonian $H$ satisfies:

$$E(\vec{\theta}) = \langle \psi(\vec{\theta}) | H | \psi(\vec{\theta}) \rangle \geq E_0 \ldots (7)$$

where $E_0$ is the true ground state energy of $H$. The objective of VQE is to minimize $E(\vec{\theta})$ over the parameter space $\vec{\theta}$, thereby approximating both the ground state and its energy corresponding to the material with the minimum dose. The following steps are followed in the VQE algorithm to get the optimized material.

**Ansatz Selection:** In order to get the minimized value of $E(\vec{\theta})$ from the equation (7), we need to explore the Hilbert space efficiently; therefore the $TwoLocal$ Ansatz is used, which is provided by Qiskit. Figure 5 shows the ansatz consisting of alternating layers of $CZ$ entangling gates and $R_y$ single-qubit rotations used to construct the quantum state $|\psi(\vec{\theta})\rangle$. It is well-suited for IBM's superconducting qubit architectures. It also offers a good balance between expressivity and hardware efficiency [19]. For each problem instance (depth), the number of qubits equals the number of candidate materials, and the ansatz is constructed accordingly.

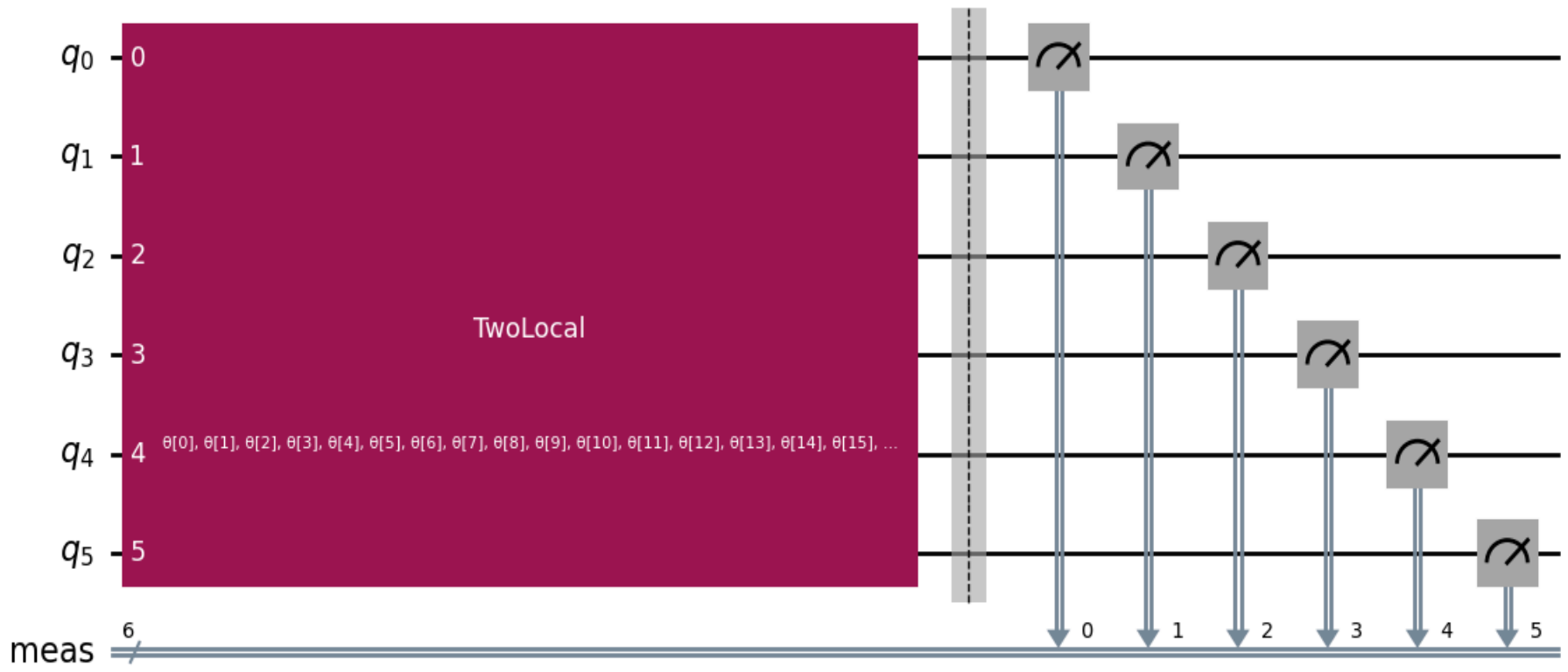

*Fig. 5 Quantum circuit representation of VQE Ansatz for QUBO-Based Material Selection Optimization for GCR as well as SPE.*

**Classical Optimization**

To iteratively tune the ansatz parameters, the COBYLA (Constrained Optimization BY Linear Approximations) optimizer [20] is employed. It updates the parameter vector $\vec{\theta}$ in a stepwise manner to minimize the expected energy of the Hamiltonian. COBYLA is selected due to its noise robustness and suitability for small-parameter optimization problems common in NISQ devices.

**Quantum Expectation Evaluation**

Once a quantum state $|\psi(\vec{\theta})\rangle$ is prepared using the ansatz for a given set of parameters ($\vec{\theta}$), the expectation value $\langle\psi(\vec{\theta})|H|\psi(\vec{\theta})\rangle$ of the Ising Hamiltonian is estimated. For this work, we utilize the Qiskit Sampler primitive, which efficiently samples bitstrings to estimate expectation values of the Hamiltonian.

**Iterative Optimization Loop**

The above steps are repeated until convergence, defined by a sufficiently small change in energy or a maximum number of iterations. The optimal parameter set $\vec{\theta}^*$ yields the quantum state $|\psi(\vec{\theta}^*)\rangle$ that best approximates the ground state of the Ising Hamiltonian for the corresponding depth.

**Measurement and Solution Decoding**

Upon convergence of the classical optimizer, the quantum circuit is measured (figure 5), and the most probable bitstring (corresponding to the lowest energy state) is identified. In this encoding, each bit directly represents the selection status of a material at the given depth (1 = selected, 0 = not selected). This process is independently repeated for every depth point, and the resulting optimal material selections are compiled to construct the complete depth-dependent material profile.

### 2.3.5 Quantum Approximate Optimization Algorithm (QAOA)

Continuing from VQE, the Quantum Approximate Optimization Algorithm (QAOA) [7] is also examined as a variational approach for the same optimization problem for a comparative analysis of these two algorithms. QAOA is particularly suited for combinatorial optimization problems and offers a different ansatz structure than VQE that alternates between two quantum operators; one derived from the problem Hamiltonian and another from a mixing Hamiltonian, parameterized by angles $\gamma$ and $\beta$. The goal, similar to VQE, is to optimize these parameters to minimize the expected energy of the system, which corresponds to the material with the minimum dose.

To implement this, a QAOA circuit is built with three layers or repetitions (reps) as shown in figure 6, where each layer applies a problem unitary $U(C, \gamma)$ and a mixing unitary $U(B, \beta)$ in sequence, as defined in equations (8) and (9), respectively. The problem unitary $U(C, \gamma)$ is expressed as,

$$U_C(\gamma) = e^{-i\gamma H_C} \ldots (8)$$

Equation (8) applies a quantum evolution based on the cost Hamiltonian $H_C$ (which encodes the QUBO objective function), where $\gamma$ is a tunable parameter representing the evolution time under this Hamiltonian. This operation biases the quantum state toward low-cost (i.e., low-dose) configurations. The mixer unitary is then employed to explore the full solution space and is given by equation (9),

$$U_M(\beta) = e^{-i\beta H_M} \ldots (9)$$

This equation (9) applies a quantum evolution based on the mixer Hamiltonian $H_M$, typically chosen as the sum of Pauli-X operators (i.e., $H_M = \sum_i X_i$), which induces transitions between computational basis states. The parameter $\beta$ controls the extent of these transitions, allowing exploration of the solution space. These gates act on the initial quantum state, typically the uniform superposition $|+\rangle^{\otimes n}$ (i.e., each qubit initialized in the $|0\rangle = \frac{1}{\sqrt{2}}(|0\rangle + |1\rangle)$ state), transforming it into a variational state that encodes high-probability solutions to the encoded material selection problem in the space radiation environment (GCR, SPE).

For this work, we use Qiskit's built-in QAOA class [21], which internally converts the QUBO matrix into a cost Hamiltonian and constructs the corresponding quantum circuit (figure 6). The COBYLA optimizer is employed to tune the variational parameters due to its robustness in low-dimensional, derivative-free optimization problems. After executing the circuit, the quantum state is measured repeatedly to obtain a distribution over bitstrings, and the most frequent bitstring is identified as the approximate solution (i.e., material with the lowest equivalent dose) to the QUBO problem for each depth. This solution is compared against classical brute force and VQE outputs to evaluate its performance in terms of both accuracy and computational efficiency.

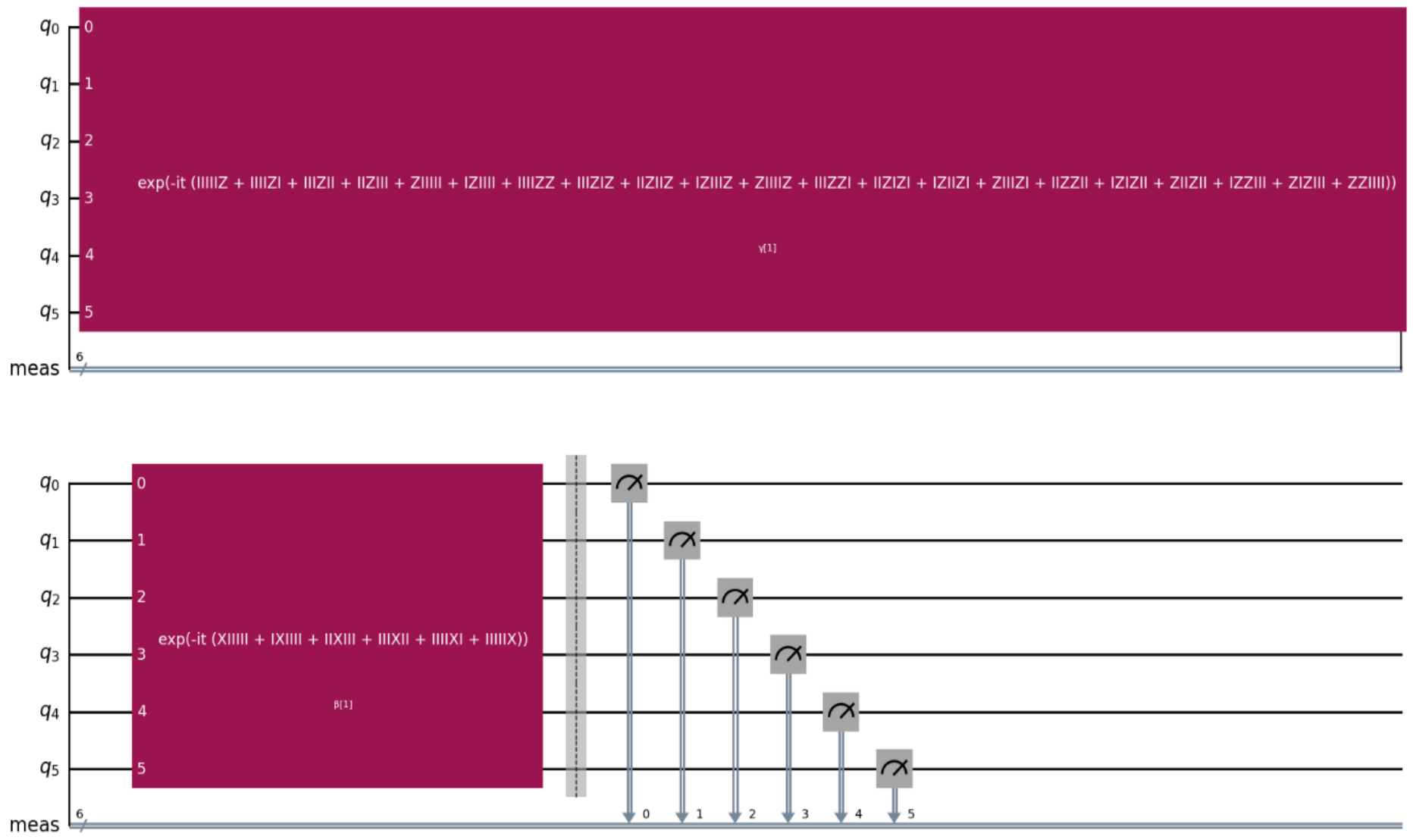

***Fig. 6*** *A single-layer QAOA circuit illustrating one cost and one mixer layer followed by measurement. Each qubit corresponds to one candidate material at a given depth.*

### 2.3.6 Classical Brute Force Method

After getting the results from VQE (section 2.3.4) & QAOA (section 2.3.5), we used the Classical Brute Force method [22] to validate and benchmark these results. This approach systematically evaluates all possible configurations of binary variables to identify the optimal solution to a Quadratic Unconstrained Binary Optimization (QUBO) material selection problem for space radiation environment. As it explores the complete solution space, this method guarantees the identification of the globally optimal configuration (i.e., the material with the lowest dose at each depth) at the cost of exponential time complexity. Specifically, for a problem with $n$ binary variables, the brute-force method evaluates all $2^n$possible bitstrings $x \in \{0, 1\}^n$. In order to find the optimal solution, we need to minimize the cost function as expressed in equation (10). For each candidate bitstring $x$, the cost is computed as:

$$\text{cost(x)} = x^T Q x + c^T x + d \ldots (10)$$

where $Q$ is a symmetric matrix of quadratic coefficients, $c$ is a vector of linear coefficients, and $d$ is an offset constant. The configuration with the lowest cost is selected as the optimal solution. While this method guarantees the exact solution, it becomes infeasible for larger problem sizes due to its exponential time complexity [23].

### 2.3.7 Qiskit Aer Simulator

The quantum algorithms discussed in Sections 2.3.4 (VQE) and 2.3.5 (QAOA) were executed using IBM's Qiskit Aer Simulator [9]. This simulator provides a high-performance, software-based quantum backend capable of emulating quantum circuits under idealized and noise-free conditions. In this study, Aer served as the primary execution environment for running and evaluating the QUBO-based quantum optimization circuits.

Although the Aer framework supports the inclusion of hardware-specific noise models [24] and imperfections, we utilized it in its default, noiseless configuration. This choice enabled an isolated assessment of the intrinsic algorithmic performance of VQE and QAOA, independent of decoherence, gate errors, or measurement noise. The resulting simulations thus represent the ideal behavior of these quantum algorithms on perfect quantum hardware. By comparing these results to those obtained via the classical brute-force method (Section 2.3.6), we benchmarked the accuracy and solution quality of the quantum approaches. Additionally, the use of Aer facilitated rapid prototyping and circuit iteration, providing valuable insights into the performance and scalability of our quantum optimization strategy for radiation shielding material selection.

## 3. RESULTS

In this section, the variation of dose equivalent and proton dose equivalent with depth are analyzed using OLTARIS and python programming. Additionally, particle-wise flux distributions are also examined to understand radiation absorption behavior and energy spectrum across different shielding materials.

Instead of absorbed dose, dose equivalent is considered because absorbed dose quantifies the energy deposited by ionizing radiation per unit mass of the tissue, whereas dose equivalent considers biological effect on tissue due to the specific type of radiation.

### 3.1 Dose Equivalent across Variable Shield Thicknesses in Different Space Radiation Environments

The dose equivalent of various materials is calculated using OLTARIS in GCR as well as in SPE environments. The results are presented in figure 7. For simulating GCR environment, the solar modulation parameter $\Phi$ is chosen as 400 MV. This represents solar minima. The solar activity follows a cycle of 11 years ranging from solar maxima to solar minima. When solar activity is maximum, the GCRs will get deflected due to strong solar magnetic field. Similarly during solar minima, the GCRs are least affected and hence have maximum intensity. This effect is known as solar modulation. To simulate the SPE environment, the flux distribution of the October 1989 event is selected in the OLTARIS simulation. It refers to one of the most intense SPE ever recorded. It produced a proton fluence exceeding $10^{10}$ protons/cm$^2$ for energies above 30 MeV, posing lethal risks to unshielded astronauts.

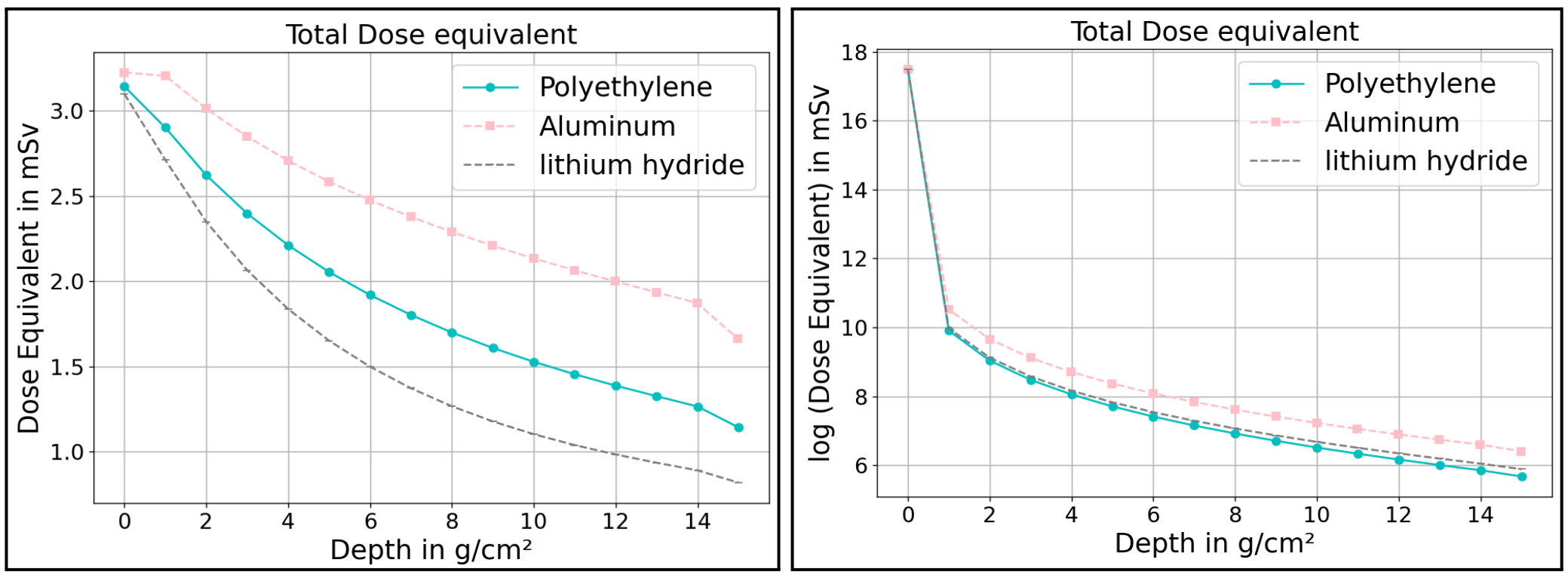

***Fig. 7.*** *Variation of dose equivalent with thickness of Aluminum, Polyethylene, and Lithium hydride in GCR at $\Phi$=400 (left) and in SPE (solar event oct 1989) (right) using OLTARIS simulation.*

It can be observed from the figure 7 (left) that the dose equivalent of lithium hydride is less than that of polyethylene and aluminum in case of the GCR environment. Also, from figure 7 (right), it can be seen that the dose equivalent of polyethylene is less than lithium hydride and aluminum in case of the SPE environment. It is also noteworthy that the dose equivalent in case of SPE is much higher than the dose equivalent from GCRs. This is due to the high intensity of SPEs, when compared to GCRs. The higher values of dose equivalent also draws attention to the need for better shielding solutions tailored to the SPE environment.

### 3.2 Proton Dose Equivalent across Variable Shield Thicknesses in different Space Radiation Environments

Protons are the most abundant particles in both GCR and SPE environments. Therefore it is necessary to understand the variation of proton dose equivalent with depth inside the shield. The proton dose equivalent in lithium hydride, polyethylene, and aluminium in the GCR and SPE environments is computed at each depth and the results are presented in figure 8.

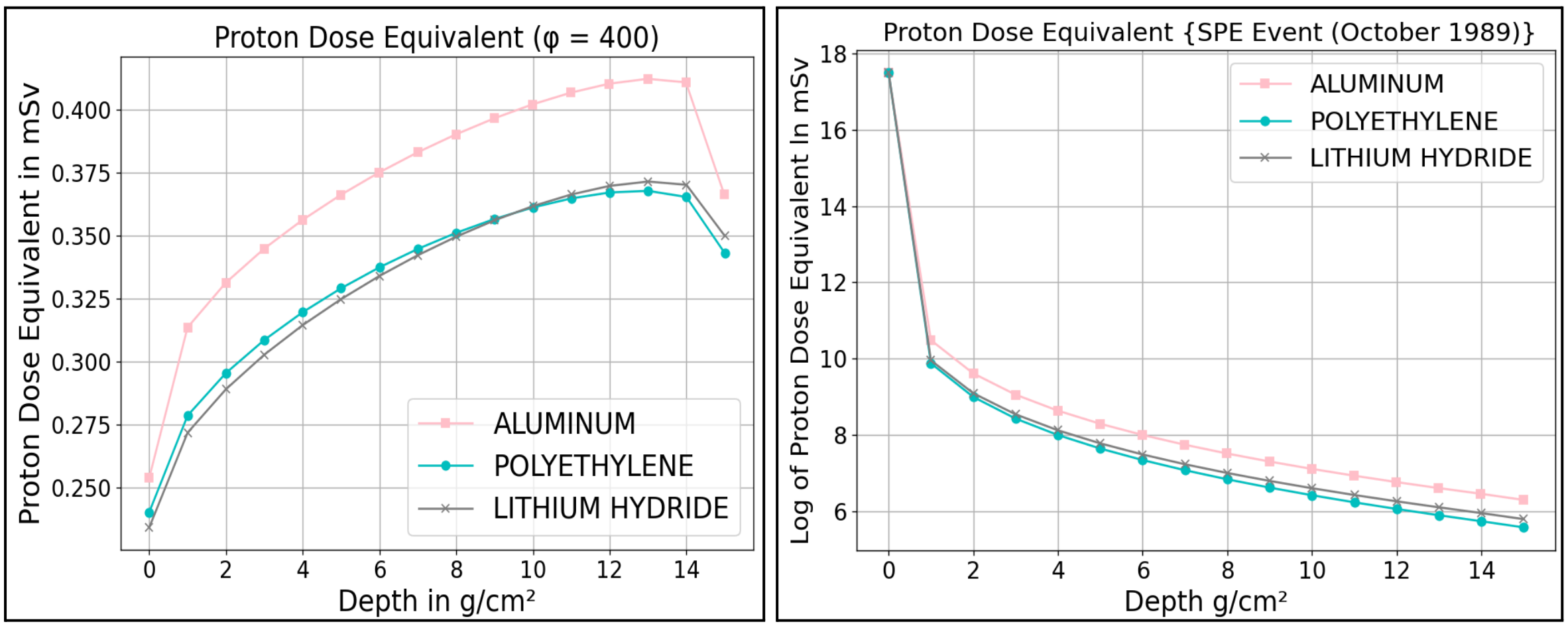

***Fig. 8.*** *Variation of proton dose equivalent with depth of Aluminum, Polyethylene and Lithium hydride in GCR at Φ = 400 (left) and in SPE (solar event oct 1989) (right) using OLTARIS simulation.*

It can be seen from the figure 8 (left) that the proton dose equivalent of Lithium hydride is less than Polyethylene and Aluminum till 10 g/cm$^2$ thickness but after 10 g/cm$^2$ Polyethylene gives less proton dose equivalent. From figure 8 (right) it can be seen that the proton dose equivalent of Polyethylene is less than Aluminum and Lithium hydride. This is in accordance with the variation of total dose equivalent in the SPE environment (figure. 7 (right)). Therefore it can be concluded that it is the effect of interaction of protons with materials that dominates in the SPE environment. In the case of GCR environment, the presence of heavy ions in the energy spectra overshadows the impact of protons.

### 3.3 Effect of Solar Modulation (ϕ) on Dose Equivalent in the GCR Environment.

The solar modulation parameter reflects the level of solar activity; higher values of ϕ indicate stronger solar activity (solar maxima), while lower values correspond to weaker activity (solar minima). To check for the consistency of the previously presented results (section 3.1) for different levels of solar modulation, the dose equivalent is computed across Aluminum, Polyethylene and Lithium hydride for different values of ϕ ranging from 400 MV (solar minimum) to 1400 MV (solar maximum). Here, the thickness of each shield is 15 g/cm$^2$. The results are shown in figure 9. During period of high solar activity, the Sun's magnetic field becomes stronger and deflects more GCRs, leading to a lower radiation dose. Conversely, at low solar activity, more cosmic rays penetrate the inner solar system, increasing the radiation dose. Hence the value of dose equivalent decreases as the value of solar modulation parameter increases. It can be observed that the relative shielding effectiveness of all the materials are consistent across all values of ϕ.

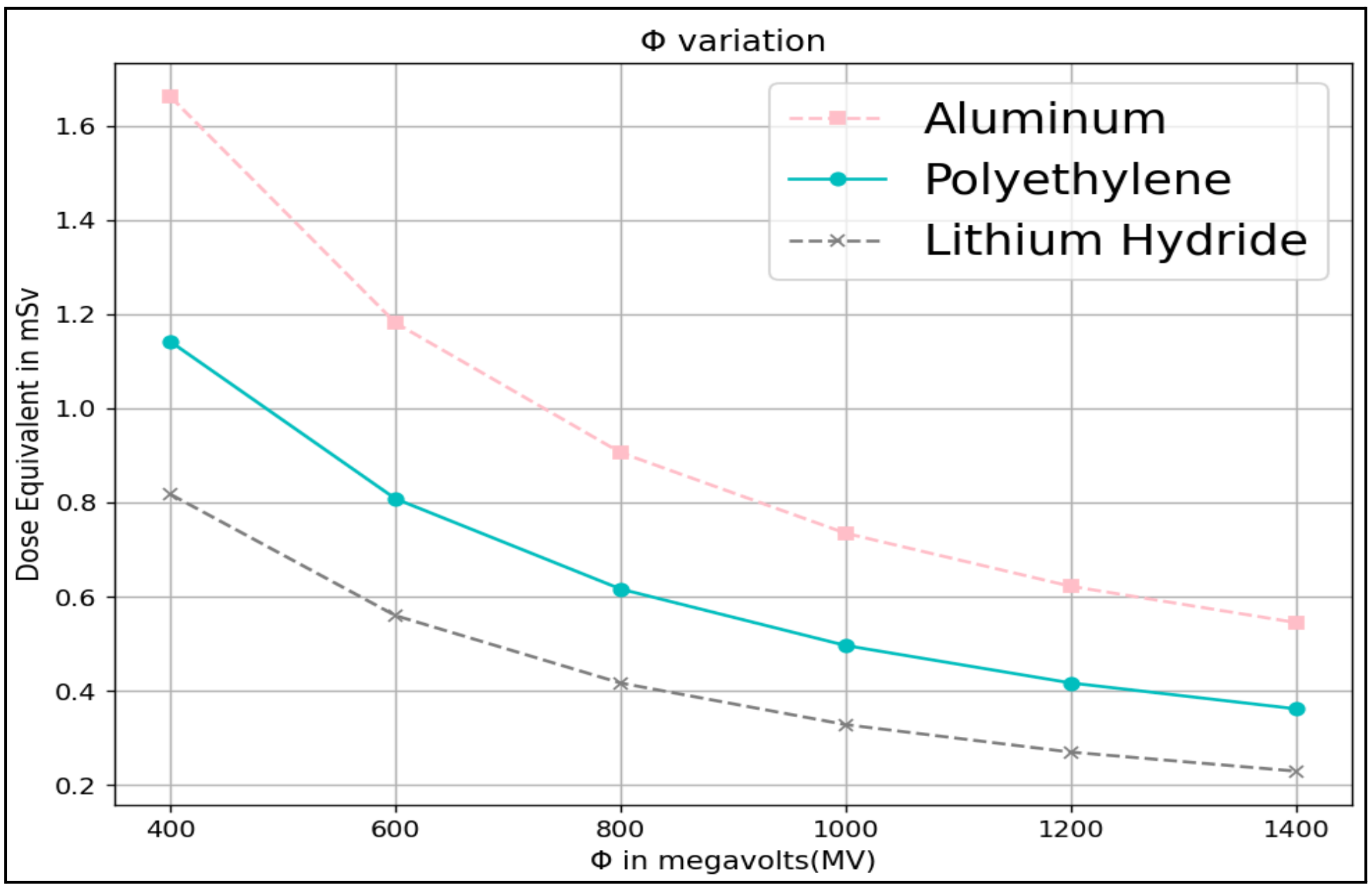

*Fig. 9. Dose equivalent variation with solar modulation in the GCR environment.*

**3.4 Comparison of Particle wise Radiation Flux in the GCR Environment**

From section 3.1 and 3.2, it was seen that the least dose equivalent from protons was produced by Polyethylene, but still it lags behind Lithium hydride in terms of total dose equivalent in the GCR environment. To delve deep into how effectively this material shields against different particles, the radiation flux before transport is compared with the radiation flux after transport for protons and iron nuclei in the GCR environment. The results are shown in figure 10. The radiation flux distribution across kinetic energy, before and after transport through a 15 g/cm $^2$ thick shield of Polyethylene, is studied for protons (left) and iron nuclei (right). An effective radiation shield is expected to reduce the incident radiation flux significantly after passing through it.

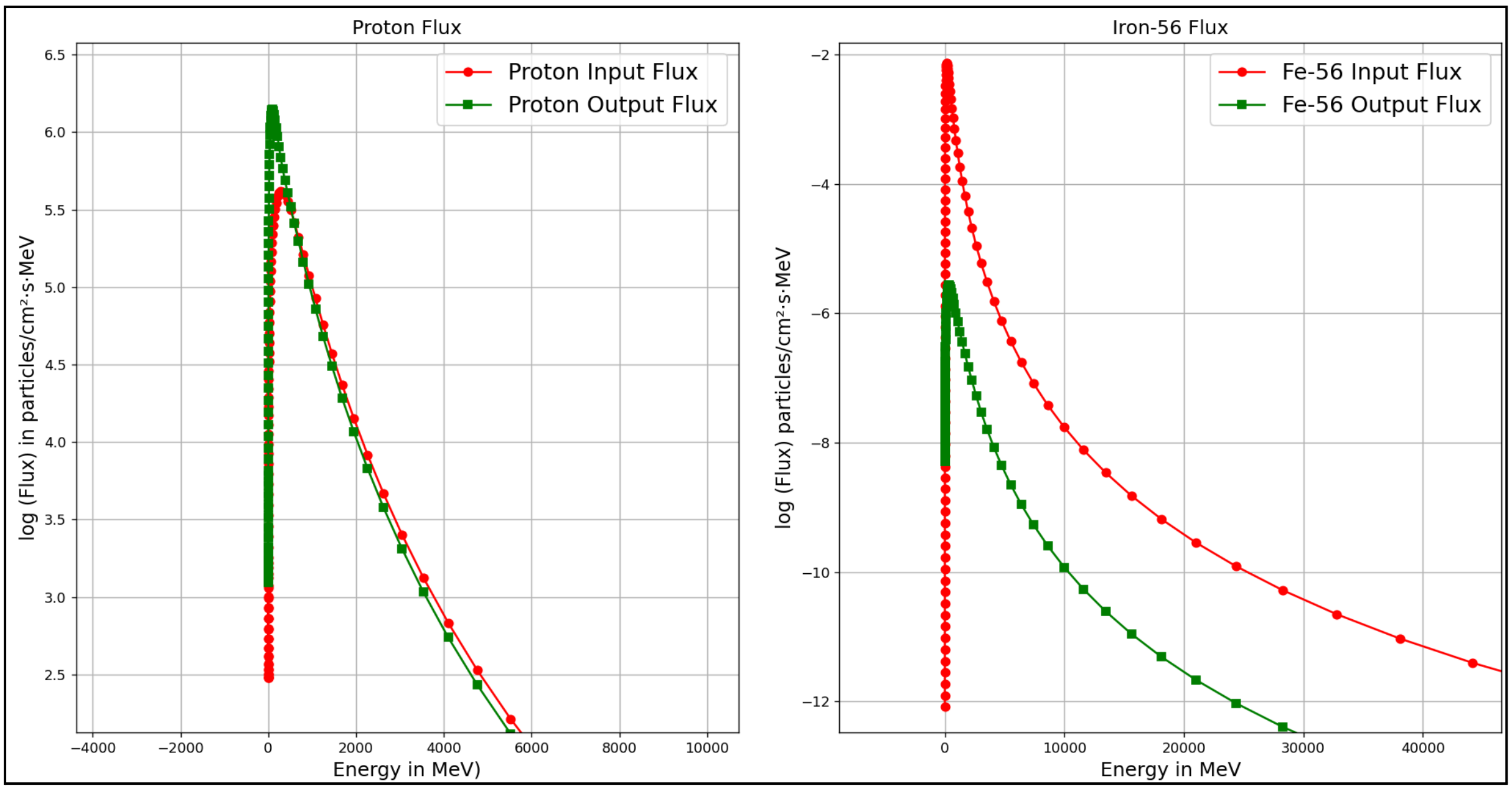

***Fig. 10*** *Flux Variation (log scale) with Kinetic Energy for Polyethylene under GCR (Φ = 400, Solar Minimum) for Protons (left) and Iron nuclei (right) using OLTARIS simulation.*

It is observed that for protons, the radiation flux after transport is greater than the flux before transport at lower energies. This implies the accumulation of protons as secondary particles from various interactions within the shield. For iron nuclei, the radiation flux after transport is considerably less than that before transport, showing that Polyethylene effectively shields against heavy ions.

In the GCR environment, the interaction of heavy ions with shielding material would generate a large number of protons as secondary particles. A material with low Z value would minimize the nuclear fragmentation of heavy ions and hence the dose equivalent. Lithium hydride has less effective atomic number than Polyethylene (Z of Li = 3 & Z of C = 6) and hence it performs better in the GCR environment (section 3.1).

### 3.5 Comparison of Particle wise Radiation Flux in the SPE Environment

The flux distribution of various particles needs to be evaluated in the SPE environment as well. The radiation flux of protons and iron nuclei before and after transport through a 15 g/cm$^2$ Polyethylene shield in the SPE environment is analyzed. The results are shown in figure 11.

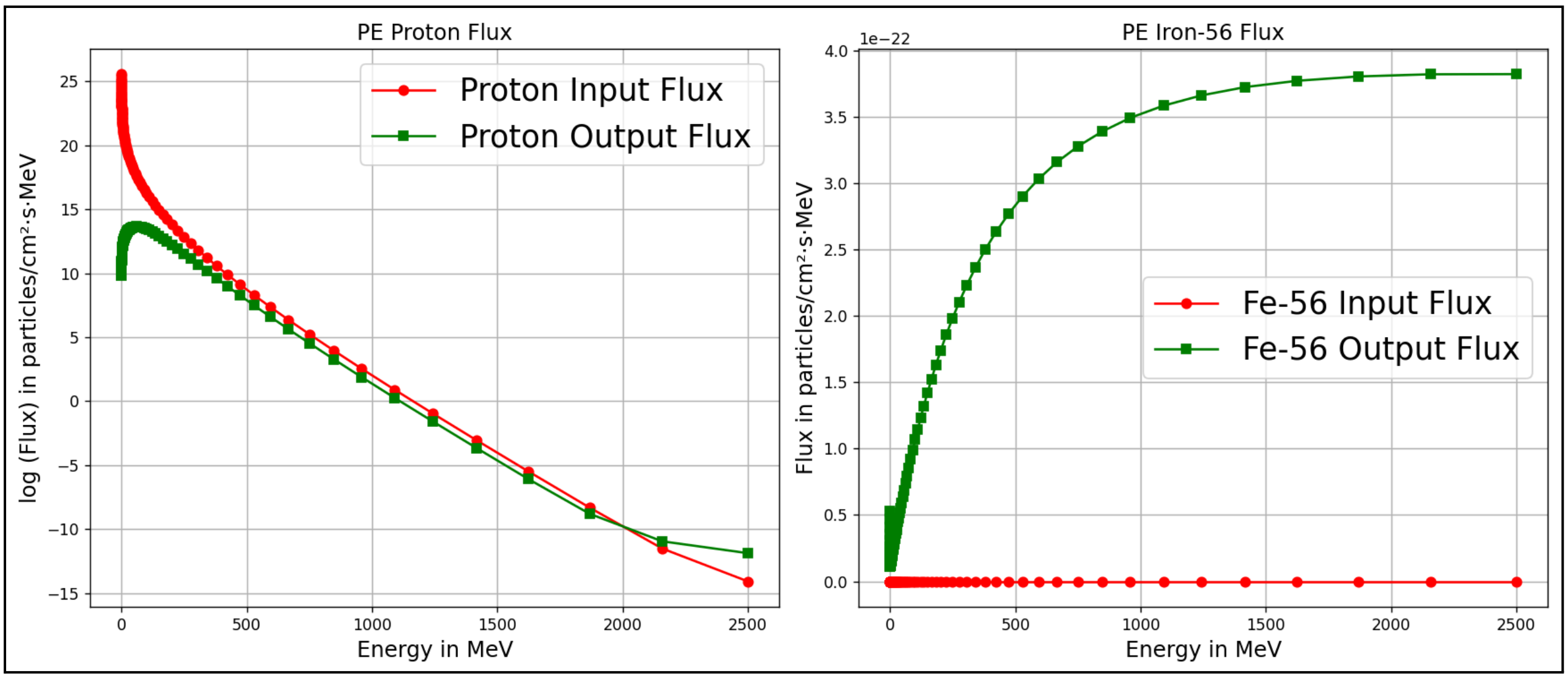

***Fig. 11*** *Flux Variation (log scale) with Kinetic Energy for Polyethylene under SPE Oct 1989 for Protons (left) and Iron nuclei (right) using OLTARIS simulation.*

In the case of protons (figure 11 (left)), it is observed that the radiation flux after transport is lesser than the flux before transport except at energies greater than 2 GeV. On the other hand, for iron nuclei (figure 11 (right)), the radiation flux after transport is negligible (of the order of $10^{-22}$) and the flux before transport is zero as heavy ions like iron are not present in the SPE spectrum. It suggests that heavy ions are being produced as secondary particles but not in a considerable amount. In the case of a SPE, as heavy ions are absent in input spectra, the ability to absorb protons alone decides the shielding effectiveness of materials (section 3.2).

### 3.6 High Hydrogen content Shielding Materials for SPE Environment

Protons lose energy within the radiation shield mainly through elastic and inelastic scattering. Hydrogen atoms within the shielding material interact with incoming protons through elastic scattering and maximum energy is transferred in this interaction. Among the materials considered, Polyethylene has a greater mass fraction of hydrogen than Lithium hydride. From section 3.1 and 3.2, it was seen that Polyethylene performed better than Lithium hydride in the SPE environment. To understand whether the greater hydrogen content of Polyethylene is the reason for its better performance under intense proton exposure, the dose equivalent of few materials (Lithium borohydride, Ammonia borane and Beryllium hydride) with even greater mass fraction of hydrogen is evaluated and analysed in the SPE environment. The results are shown in figure 12.

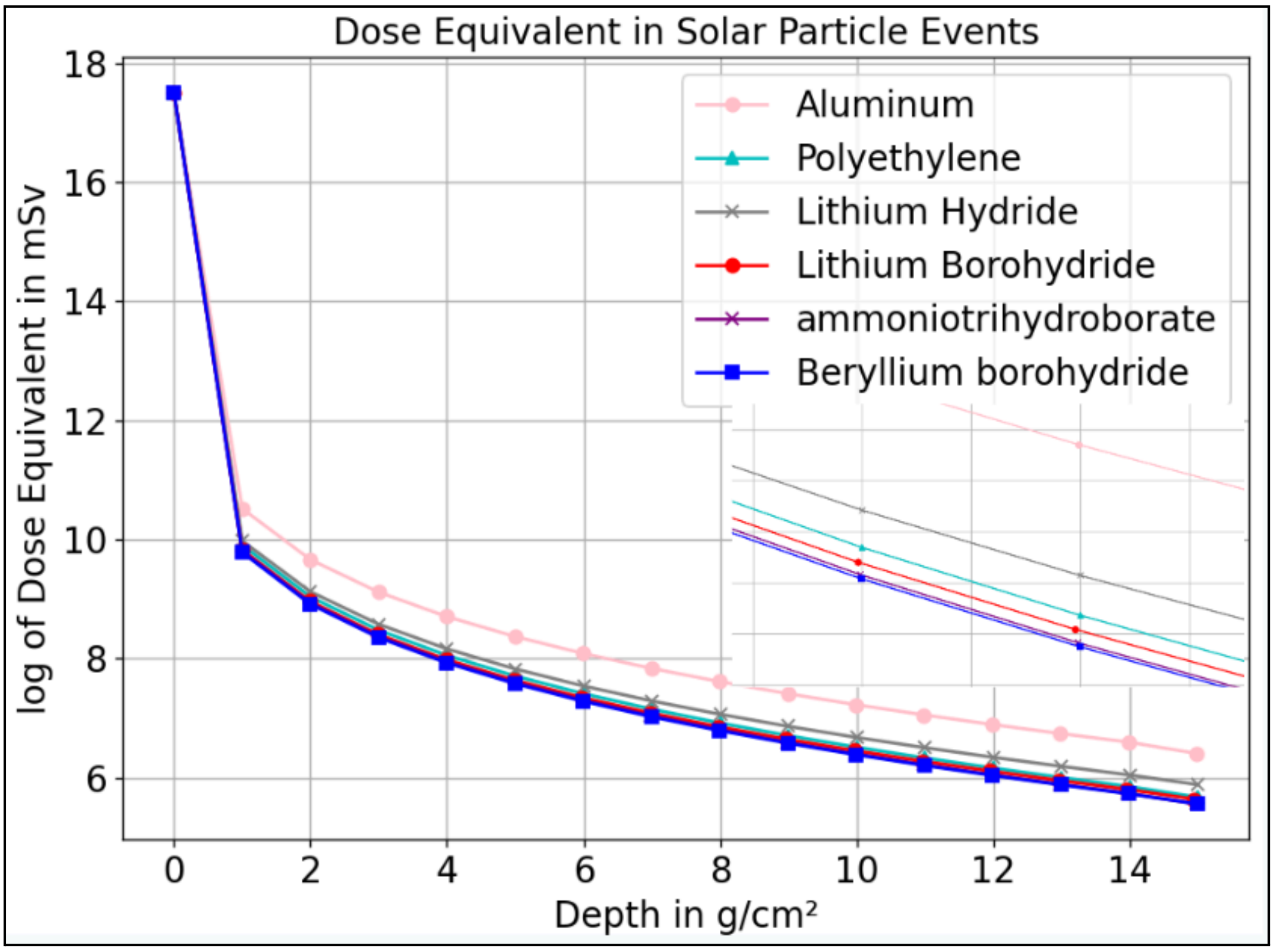

***Fig. 12*** *Variation of Dose Equivalent (log scale) with Shield thickness for Lithium hydride, Polyethylene, Aluminium, Lithium borohydride, Ammonia Borane and Beryllium borohydride [25] using OLTARIS simulation in solar event oct 1989.*

It can be seen from figure 12 that the dose equivalent of Beryllium borohydride, which has the highest hydrogen mass fraction, is less than the dose equivalent of other listed materials. Hence this concludes that high hydrogen content materials give a lesser dose equivalent in the case of Solar Particle Events.

### 3.7 Proton Dose Equivalent in High hydrogen content Shielding Materials for the SPE Environment

In this study dose equivalent from protons is compared in various high hydrogen containing materials as well as in Aluminum. The results are presented in figure 13.

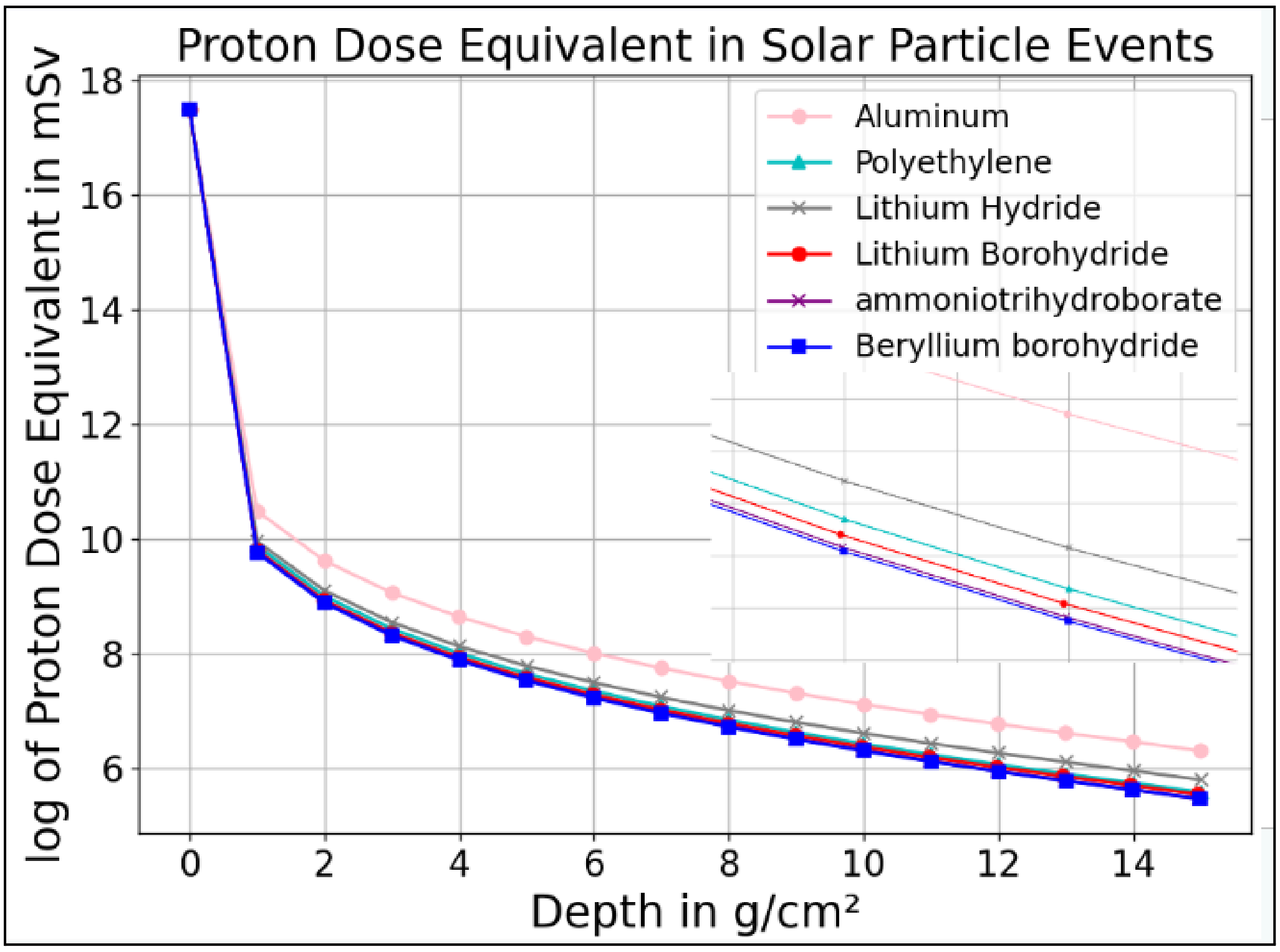

***Fig. 13*** *Variation of Proton Dose Equivalent (log scale) with Shield thickness for Lithium hydride, Polyethylene, Aluminium, Lithium borohydride, Ammonia borane and Beryllium borohydride [25] using OLTARIS simulation in solar event oct 1989.*

It can be seen from the figure that the proton dose equivalent of Beryllium borohydride is less than that of Lithium hydride, Polyethylene, Aluminium, Lithium borohydride and Ammonia borane. This result is consistent with the observation in section 3.6.
These results put forward an important aspect that needs to be considered while developing shielding solutions for SPEs. Materials like Beryllium borohydride can have potential applications in the construction of storm shelters and specialised suits for future space missions.

### 3.8 Quantum Optimization of Shielding Materials in different Space Radiation Environments

The detailed analysis in the previous sections highlighted how radiation shielding effectiveness varies significantly across materials, environments (GCR & SPE), and depth. As more high-hydrogen materials are introduced and parameters such as dose equivalent, proton flux, and secondary particle production are considered, the material selection process becomes increasingly complex and multi-dimensional. Rather than manually evaluating every possible configuration, an optimization framework is essential to efficiently identify the best shielding material.

To address this, we mapped the OLTARIS-generated dose equivalent data into a Quantum optimization Model i.e. QUBO model, allowing us to use quantum algorithms, specifically QAOA and VQE, to perform material selection. These algorithms were implemented using both the Aer simulator (Quantum Simulator) and classical brute-force methods for benchmarking, enabling a direct comparison between quantum and classical approaches for solving shielding optimization problems in both GCR and SPE environments. Their different parameters, including circuit depth, number of qubits used, time complexity, and optimization values vs depth plots have been compared in the following sub-sections.

### 3.8.1 Comparison of Quantum and Classical Optimization of shielding materials in GCR environment

Table 2 shows the comparison of various parameters between Aer Simulator (VQE & QAOA) and the Classical Brute Force method for the GCR environment. Here, $p$ = ansatz depth, $n_q$ = number of qubits, $n_c$ = total number of classical binary variables $iter$ = maximum number of iterations. For each evaluated depth, all the methods, classical brute force, and VQE & QAOA on Aer simulator, identified LiH (Lithium Hydride) as the optimal shielding material for Galactic Cosmic Rays (GCRs). This uniformity across approaches strongly validates the accuracy of the VQE & QAOA implementation [26]. Further, this complements the results from the OLTARIS simulation as shown in Fig. 7, which confirms the feasibility study of using quantum algorithms (VQE & QAOA) for space radiation shielding applications.

| Variables | Aer Simulator (VQE) | Aer Simulator (QAOA) | Classical Brute Force |
|---|---|---|---|
| Qubits | 3 | 3 | - |
| Circuit Depth | 3 | 3 | - |
| Minimum Dose (per Depth) | Same as Brute Force | Same as Brute Force | Exact (LiH gives minimum) |
| Best Material (per depth) | Same as Brute Force | Same as Brute Force | LiH (Lithium Hydride) |
| Time Complexity | $\mathcal{O}(p \cdot n_q \cdot \text{iter})$ | $\mathcal{O}(2^{n_q} \cdot p \cdot \text{iter})$ | $\mathcal{O}2^{n_c}$ |
| Average Time Taken (per depth) | ~ 0.210 seconds | ~ 0.038 seconds | ~ 0.000135 seconds |

***Table 2**: Comparison of various parameters between Aer Simulator (VQE & QAOA) and Classical Brute Force method for GCR environment.*

Notably, the time performance showed variation: the Classical Brute Force is the fastest (~0.000135 s), followed by QAOA (~0.038 s), while VQE took ~0.210 seconds. This increased duration of the quantum algorithms is primarily attributed to several factors: Circuit depth for expressivity, number of iterations for optimal results, execution overhead from circuit transpilation and error mitigation procedures, and repeated sampling (shots) needed to obtain reliable statistics. Despite the overhead and large number of shots, the convergence to the correct solution, that is LiH, gave the minimum dose in all three configurations, which highlights VQE's & QAOA's promise for practical small-scale optimization problems for space applications. To improve the execution time, several strategies can be employed: reducing circuit depth through optimized ansatz design, minimizing the number of measurement shots using adaptive sampling techniques, and leveraging more efficient transpilation strategies tailored to specific optimization problems for shielding materials.

To better understand the behavior and consistency of the VQE & QAOA-based optimization approach, it is important to examine how the algorithms perform across varying depths. Fig. 14 illustrates the comparison of optimization values obtained from VQE as well as QAOA, simulated using the Qiskit Aer simulator as a function of shielding depth of materials provided from the output of the OLTARIS simulation. Each point represents the minimum energy corresponding to the optimized LiH material obtained by the methods for a single instance of the QUBO-encoded material selection problem at a particular depth.

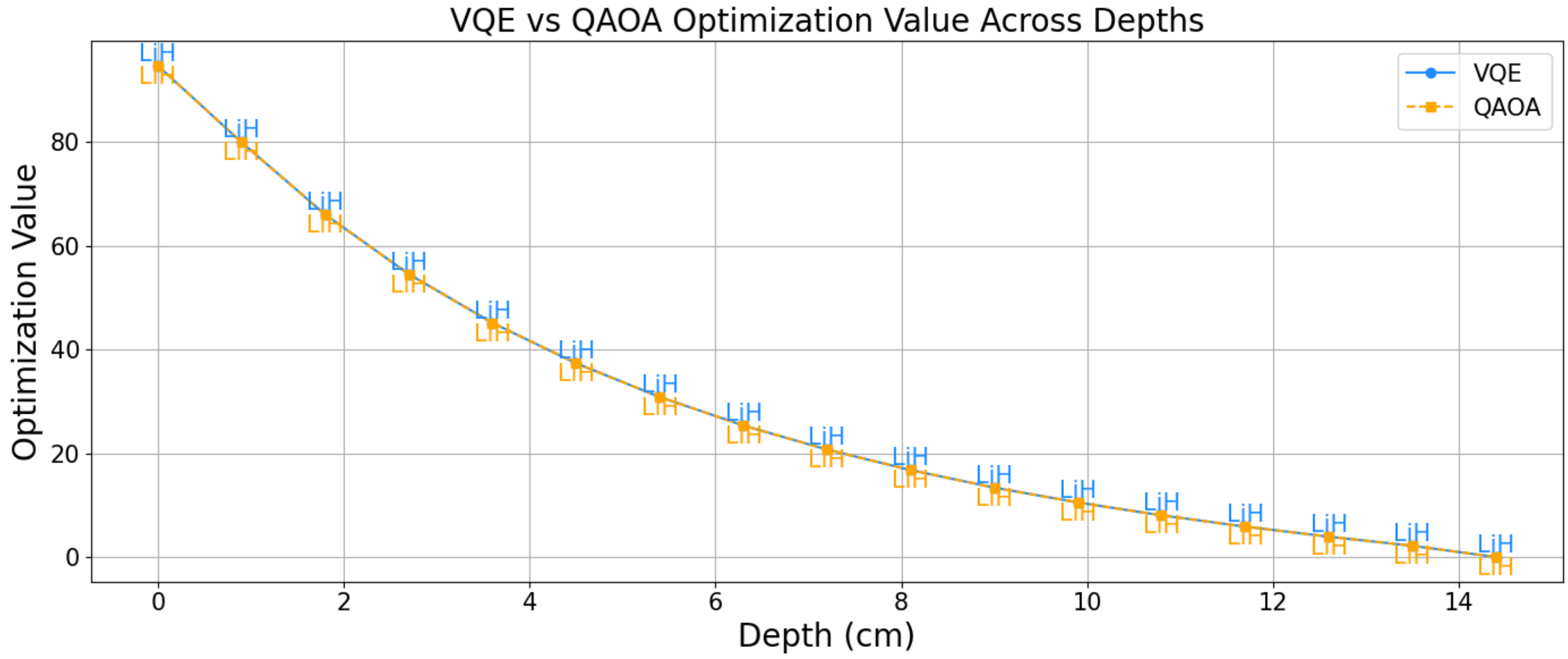

***Fig.14*** *Comparison of Optimization Values of VQE & QAOA vs Depth (cm)(GCR)*

The trend of monotonically decreasing optimization values with increasing depth indicates that the optimal solutions correspond to lower absorbed doses at greater depths, which aligns with physical intuition: as shielding thickness increases, radiation penetration decreases. The smooth downward curve also suggests a diminishing return on dose reduction as depth increases; the marginal benefit (reflected by the change in optimization value) gradually decreases. This trend highlights a natural saturation in shielding effectiveness.

Moreover, the absence of noise and the stability of the curve confirm that they converged consistently across all depths under ideal (noiseless) simulation conditions. This validates both the numerical stability of the algorithm and the correctness of the QUBO-to-Hamiltonian mapping for this material optimization problem.

### 3.8.2 Comparison of Quantum and Classical Optimization of shielding materials in SPE environment

To assess the applicability of hybrid quantum algorithms in the Solar Particle Event (SPE) environment, we extended our QUBO-based optimization approach using dose equivalent data from OLTARIS. The QUBO problem was solved using VQE and QAOA on the Qiskit Aer simulator, and the results were benchmarked against classical brute-force optimization. *Table 3* summarizes the key parameters across all three methods. Both VQE and QAOA, using six qubits and a circuit depth of three, consistently matched the classical brute-force method in identifying the best shielding material and corresponding minimum dose equivalent values across all depths. As observed in the GCR case, classical brute force was the fastest (~0.00029 s), followed by QAOA (~0.079 s) and VQE (~0.329 s), with time complexity and overheads scaling with problem size and circuit depth.

| **Variables** | **Aer Simulator (VQE)** | **Aer Simulator (QAOA)** | **Classical Brute Force** |
|---|---|---|---|
| Qubits | 6 | 6 | - |
| Circuit Depth | 3 | 3 | - |
| Minimum Dose (per Depth) | Same as Brute Force | Same as Brute Force | See *figure11* below |
| Best Material (per depth) | Same as Brute Force | Same as Brute Force | See *figure11* below |
| Time Complexity | $\mathcal{O}(p \cdot n_q \cdot \text{iter})$ | $\mathcal{O}(2^{n_q} \cdot p \cdot \text{iter})$ | $\mathcal{O}2^{n_c}$ |
| Average Time Taken (per depth) | ~ 0.329 seconds | ~ 0.079 seconds | ~ 0.00029 seconds |

***Table 3:*** *Comparison of various parameters between Aer Simulator (VQE & QAOA) and Classical Brute Force method for SPE environment.*

To further examine algorithm behavior across depths, figure 15 presents the optimization values for each depth for the SPE environment. In this figure, LiB is Lithium borohydride and Boro_H is Beryllium borohydride. Both quantum solvers exhibited a steep reduction in absorbed dose from 0 - 2 cm, followed by a saturation trend beyond 4 cm. This behavior aligns with physical expectations, where increased shielding reduces radiation dose, but offers diminishing returns at larger thicknesses. The near-identical and stable curves from VQE and QAOA confirm numerical robustness and convergence under noiseless simulation conditions.

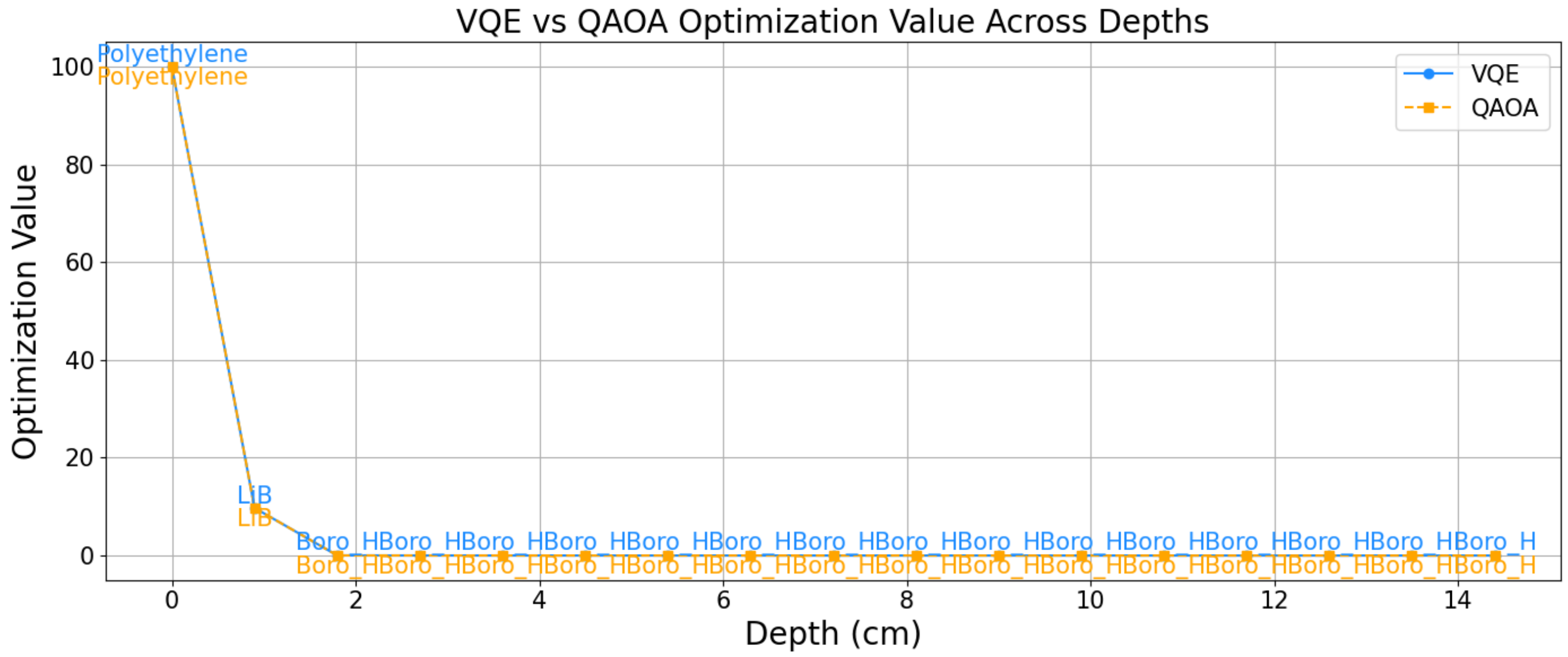

*Fig. 15 Comparison of Optimization Values of VQE & QAOA vs Depth (cm) (SPE)*

These results, in both GCR as well as SPE environments, support the idea that for certain low-dimension, high-impact optimization problems, hybrid quantum algorithms can match classical accuracy, while offering insights into future scalability when qubit counts increase [27]. A promising extension of this work is toward multi-layered shielding, where both material selection and ordering must be optimized. Though more complex and qubit-intensive, such problems are well-suited for hybrid quantum algorithms like VQE and QAOA as hardware capabilities grow.

**4. Summary:**

This study analyzed the radiation shielding effectiveness of various Metal hydrides, Polymers [28] and traditional materials in which we emphasised on their hydrogen storage capacity, proton absorbance property as a key fact. The dose equivalent was examined as a function of shield thickness, alongside we also studied the contribution of different space radiation particles such as protons and heavy ions (iron) in GCR and SPE environments.

The dose equivalent of Aluminum (conventional material), Polyethylene (polymer) and Lithium hydride (metal hydride) was studied in GCR as well as SPE environments. Polyethylene generated the least dose equivalent in the SPE environment, whereas Lithium hydride performed the best in the GCR environment. To investigate the reason for this environment dependency of shielding performance, proton dose equivalent was studied separately in both the environments. It was observed that in the SPE environment, the shielding effectiveness is directly proportional to the ability of proton attenuation of materials and materials with greater hydrogen mass fraction are better proton attenuators. This observation was verified by evaluating dose equivalent from a few additional materials with higher hydrogen content (Ammonia borane and Beryllium borohydride). Further, the effect of solar modulation in dose equivalent was evaluated. The comparison of proton and iron flux after transport with the boundary conditions revealed that in the GCR environment, the ability of materials to minimize the dose equivalent from heavy ions like iron decides the overall shielding effectiveness.

In parallel, this study also demonstrates the feasibility of using VQE & QAOA to solve QUBO problems for space applications, through a practical case study focused on depth-dependent material selection for radiation shielding, using the dose-equivalent data sourced from OLTARIS across varying depths. VQE & QAOA are executed on the Aer simulator and yielded results are found to be consistent with classical brute-force methods, identifying Lithium Hydride (LiH) as the optimal shielding material in the Galactic Cosmic Rays (GCR) environment and Beryllium borohydride in the case of the SPE environment. The results obtained from Quantum optimization also agree with the results obtained from OLTARIS simulations. This work validates our QUBO-to-Ising mapping and shows the reliability of VQE & QAOA on near-term quantum devices. The corresponding optimization trends are illustrated in figure 14 and figure 14, showing how the optimization values (representing the dose-equivalent data) associated with the most effective shielding material (i.e., the material yielding the lowest dose equivalent) at each depth change across both GCR and SPE environments. The smooth, monotonically decreasing curves demonstrate physically consistent behavior, with diminishing returns in dose reduction at greater depths. These trends not only reflect the correctness of the QUBO encoding but also validate the stability and convergence of the quantum algorithms under noiseless simulation conditions.

From a performance standpoint, the classical brute-force was the fastest, followed by the Aer simulator. VQE & QAOA are able to converge to correct solutions quickly, underscoring the promise of hybrid quantum-classical approaches for small-scale problems. The proof-of-concept circuits operated on three and six qubits per depth, demonstrating feasibility on current NISQ-era hardware. Overall, this study bridges classical radiation modeling and quantum optimization by showing how dose-equivalent data from OLTARIS can be embedded into a quantum (QUBO) framework and solved using quantum variational algorithms. It offers new insights into how various radiation particles interact with shielding materials, while establishing a viable quantum-classical pipeline for space radiation protection challenges. The impact of this work lies in advancing lightweight, high-performance shielding materials for personal shielding garments, lightweight jackets, and storm shelters designed for astronaut safety in deep space missions.

## 5. Future Scope

This current simulation study relied on OLTARIS, a deterministic radiation transport tool, future work could integrate Monte Carlo-based methods such as GEANT4 to capture more detailed physics, including stochastic secondary particle generation. Additionally, the design of multilayer shielding systems that combine different material types offers a promising direction optimizing radiation attenuation while maintaining structural and mechanical integrity. Advanced nanomaterials such as Boron Nitride Nanotubes (BNNTs) [29] warrant further exploration due to their excellent neutron absorption properties and potential to form lightweight, durable composites, when embedded in polymers. Another innovative concept involves water-filled Carbon Nanotubes; while water is a highly effective radiation shield, its volatility in space is a concern that can be mitigated by encapsulating it within nanostructured carriers. These materials were not studied in this work as the simulation tool does not consider their properties as nanomaterials effectively.

On the aspects of Quantum optimization, the current study demonstrates the feasibility of using quantum algorithms for radiation shielding optimization, where all simulations were conducted on an ideal, noiseless quantum simulator, i.e., AerSimulator, provided by IBM's Qiskit. A key direction for future work involves incorporating realistic noise models, which is essential to mimic the real quantum hardware with decoherence and gate errors to evaluate the performance and robustness of Variational Quantum Algorithms.

Moreover, deploying the quantum circuits on actual IBM Quantum hardware will allow benchmarking against physical constraints, helping to assess the viability of hybrid algorithms in real-world settings. Given the algorithm's scalability, the framework is particularly suitable for analyzing large datasets, enabling precise and fast optimization across an extensive range of material configurations. Additionally, the quantum simulation of space radiation interactions with shielding materials can be done using quantum algorithms such as VQE & QAOA, both on quantum simulator as well as IBM & D-Wave's quantum hardware, which represents a promising long-term research goal, offering the potential for end-to-end quantum-enabled design pipelines for future space missions.